\documentclass[showpacs,preprintnumbers,superscriptaddress]{revtex4}
\usepackage{CJK}
\usepackage{amsmath,amssymb,graphicx,bm}
\begin{document}
\begin{CJK*}{GBK}{song}
\title{Horizon thermodynamics in $f(R,R^{\mu\nu}R_{\mu\nu})$ theory}

\author{Haiyuan Feng}
\affiliation{College of Physical Science and Technology, Hebei University, Baoding 071002, China}
\author{Rong-Jia Yang \footnote{Corresponding author}}
\email{yangrongjia@tsinghua.org.cn}
\affiliation{College of Physical Science and Technology, Hebei University, Baoding 071002, China}
\affiliation{Hebei Key Lab of Optic-Electronic Information and Materials, Hebei University, Baoding 071002, China}
\affiliation{National-Local Joint Engineering Laboratory of New Energy Photoelectric Devices, Hebei University, Baoding 071002, China}
\affiliation{Key Laboratory of High-pricision Computation and Application of Quantum Field Theory of Hebei Province, Hebei University, Baoding 071002, China}
\begin{abstract}
We investigate whether the new horizon first law still holds in $f(R,R^{\mu\nu}R_{\mu\nu})$ theory. For this complicated theory, we first determine the entropy of black hole via Wald method, then we derive the energy by using the new horizon first law, the degenerate Legendre transformation, and the gravitational field equations. For application, we consider the quadratic-curvature gravity and firstly calculate the entropy and the energy for a static spherically symmetric black hole, which reduces to the results obtained in literatures for a Schwarzschild-(A)dS black hole.
\end{abstract}
\pacs{04.07.Dy, 04.50.Kd, 04.20.Cv}
\keywords{black hole, entropy, energy, new horizon first law}
\maketitle
\section{Introduction}
Black hole, predicted in general relativity, is an object of long-standing interest to physicists. Bekenstein firstly proposed that black holes
actually possess entropy \cite{Bekenstein:1973ur}. In sharp contrast to standard thermodynamic notions where entropy is supposed to be a function of volume, he suggested that the entropy of black hole is proportional to the horizon area. Since then a variety of different theoretical methods have been used to calculate the Bekenstein-Hawking entropy, such as starting from quantum fields near the horizon \cite{tHooft:1984kcu}, from quantum field theory in a fixed background \cite{Hawking:1974rv}, from entanglement entropy \cite{Bombelli:1986rw}, from string theory \cite{Strominger:1996sh, Lunin:2002qf, Peet:1997es, Callan:1996dv, Aharony:1999ti, Mathur:2005zp}, from loop quantum gravity \cite{Ashtekar:1997yu, Livine:2005mw}, from Noether charge \cite{Wald:1993nt, Iyer:1994ys}, from induced gravity \cite{Frolov:1997up}, from causal set theory \cite{Rideout:2006zt}, from symmetry near horizon \cite{Majhi:2011ws, Carlip:2017xne}, from inherently global characteristics of a black hole spacetime \cite{Hawking:1998jf}, etc. It has been shown that the classical Bekenstein-Hawking entropy depends not only on the black hole parameter, but also on the coupling which reduces Lorentz violation \cite{Chen2015}. These methods are complicated. Finding a simpler way to calculate the entropy of a black hole is an important task.

Energy is another important issue besides entropy in black hole physics. In higher-order gravitational theories, the energy of black hole is still an open problem. Several efforts to find a satisfactory answer to this issue have been carried out \cite{Cognola:2011nj, Deser:2002jk, Deser:2007vs, Abreu:2010sc, Cai:2009qf}. It was shown that the entropy and energy of black hole can be simultaneously obtained in Einstein's gravity via the horizon first law \cite{Padmanabhan:2002sha}, but it can not work in higher-order gravitational theories. Recently a new horizon first law, in which both the entropy and the free energy are derived concepts, was suggested in Einstein's gravity and Lovelock's gravity, the standard horizon first law can be recovered by a Legendre projection \cite{Hansen:2016gud}. In \cite{Zheng:2018fyn}, it was found that the new horizon first law still work in $f(R)$ theories by introducing the effective curvature fluid: it can give not only the energy but also the entropy of black holes, which reproduce the known results in literatures. Here we will consider the new horizon first law and the entropy and the energy issues  in $f(R,R^{\mu\nu}R_{\mu\nu})$ gravity.

The paper is organised as follows. In section II, we briefly review the new horizon first law. In section III, we discuss the entropy and the energy of black holes in $f(R,R^{\mu\nu}R_{\mu\nu})$ theory. In section IV, applications are considered. Finally, We will briefly summarize our results in section V.

\section{The new horizon first law}
According to the suggestion proposed in \cite{Yang:2014kna}, that the source of thermodynamic system is also that of gravity, the radial component of the stress-energy tensor can act as the thermodynamic pressure, $P=T^{r}_{~r}|_{r_{+}}$, then at the horizon of Schwarzschild black hole the radial Einstein equation can be written as
\begin{eqnarray}
\label{p}
P=\frac{T}{2r_{+}}-\frac{1}{8\pi r^2_{+}},
\end{eqnarray}
which can be rewritten as a horizon first law after a imaginary displacement of the horizon, $\delta E=T\delta S-P\delta V$, with $E$ the quasilocal energy and $S$ the horizon entropy of the black hole \cite{Padmanabhan:2002sha}. Since the temperature $T$ in the equation (\ref{p}) is identified from thermal quantum field theory, independent of any gravitational field equations \cite{Hansen:2016gud}, while the pressure $P$ in (\ref{p}), according to the conjecture proposed in \cite{Yang:2014kna}, is identified as the radial component of the matter stress-energy, so it is reasonable to assume that the radial field equation of a gravitational theory under consideration takes the form \cite{Hansen:2016gud}
\begin{eqnarray}
\label{p1}
P=D(r_+)+C(r_+)T,
\end{eqnarray}
where $C$ and $D$ are analytic functions of the radius of black hole, $r_+$, in general they depend on the gravitational theory under considering. Varying the equation (\ref{p}) and multiplying the geometric volume $V(r_+)$, it is straightforward to have a new horizon first law \cite{Hansen:2016gud}
\begin{eqnarray}
\label{nhfl}
\delta G=-S\delta T+V\delta P,
\end{eqnarray}
with the Gibbs free energy as
\begin{eqnarray}
\label{10}
G&=&\int^{r_+} V(r)D'(r)\,dr+T\int^{r_+} V(r)C'(r)dr \nonumber\\
&=&PV-ST-\int^{r_+} V'(r)D(r)dr,
\end{eqnarray}
and the entropy as \cite{Hansen:2016gud}
\begin{eqnarray}
\label{11}
S=\int^{r_+} V'(r)C(r)dr.
\end{eqnarray}
Under the degenerate Legendre transformation $E=G+TS-PV$, yields the energy as \cite{Zheng:2018fyn}
\begin{eqnarray}
\label{energy}
E=-\int^{r_+} V'(r)D(r)dr.
\end{eqnarray}
This procedure was firstly discussed in Einstein gravity and Lovelock gravity which only give rise to second-order field equation \cite{Hansen:2016gud}. It was generalized to $f(R)$ gravity with a static spherically symmetric black hole \cite{Zheng:2018fyn} or with a general spherically symmetric black hole \cite{Zheng:2019mvn} and was also applied to $D$-dimensional $f(R)$ theory \cite{zhu}.

In the next Section, we will investigate whether this procedure can be applied to more complicated case, such as $f(R,R^{\mu\nu}R_{\mu\nu})$ theory, whether Eqs. (\ref{11}) and (\ref{energy}) can still be used to obtain the entropy and the energy in the theory we consider.

\section{The entropy and the energy of black holes in $f(R,R^{\mu\nu}R_{\mu\nu})$ theory}
As shown in Section II, the new horizon first law works well in Einstein's theory and Lovelock gravity \cite{Hansen:2016gud} and $f(R)$ theory \cite{Zheng:2018fyn,Zheng:2019mvn,zhu}. Whether it still works in other gravitational theories, such as $f(R, R^{\mu\nu}R_{\mu\nu})$ theory? We consider this question in this Section. In four-dimensional spacetime, the general action of $f(R, R^{\mu\nu}R_{\mu\nu})$ theory with source is given by
\begin{eqnarray}
\label{17}
I=\int d^4 x\sqrt{-g}\left[\frac{f(R,R^{\mu\nu}R_{\mu\nu})}{16\pi}+L_m\right],
\end{eqnarray}
where $L_{\rm m}$ is the matter Lagrangian and $f(R, R^{\mu\nu}R_{\mu\nu})$ is a general function of the Ricci scalar $R$ and the square of the Ricci tensor $R_{\mu\nu}$. We take  the units $G=c=\hbar=1$. Varying the action (\ref{17}) with respect to metric $g_{\mu\nu}$, yields the gravitational field equations as
\begin{eqnarray}
\label{18}
G_{\mu\nu}&\equiv &
R_{\mu\nu}-\frac{1}{2}Rg{_{\mu\nu}}=8\pi\left[\frac{T_{\mu\nu}}{f_{R}}+\frac{1}{8\pi}\Omega_{\mu\nu}\right],
\end{eqnarray}
where $f_{R}\equiv\frac{\partial f}{\partial R}$ and $T_{\mu\nu}=\frac{2}{\sqrt{-g}}\frac{\delta L_{m}}{\delta g^{\mu\nu}}$ is the energy-momentum tensor of matter. $\Omega_{\mu\nu}$ is the tress-energy tensor of the effective curvature fluid and is given by
\begin{eqnarray}
\label{19}
\Omega_{\mu\nu}=\frac{1}{f_{R}}\left[\frac{1}{2}g_{\mu\nu}(f-Rf_{R})+\nabla_{\mu}\nabla_{\nu}f_{R}-g_{\mu\nu}
\Box f_{R}-2f_{X}R^{^{\alpha}}_{\mu}R_{\alpha\nu}-\Box(f_{X}R_{\mu\nu})-g_{\mu\nu}\nabla_{\alpha}\nabla_{\beta}(f_{X}R^{\alpha\beta})
+2\nabla_{\alpha}\nabla_{(\mu}(R^{\alpha}_{\nu)}f_{X})\right],
\end{eqnarray}
where $A_{(\mu\nu)}=\frac{1}{2} (A_{\mu\nu}+A_{\nu\mu}$), $X\equiv R^{\mu\nu}R_{\mu\nu}$, and $f_{X}\equiv\frac{\partial f}{\partial X}$. Thanking of the following two derivational relations
\begin{eqnarray}
\nabla_{\alpha}\nabla_{\beta}(f_{X}R^{\alpha\beta})=R^{\alpha\beta}\nabla_{\alpha}\nabla_{\beta}f_{X}+(\nabla^{\beta}R)(\nabla_{\beta}f_{X})+\frac{1}{2}f_{X}\Box R,
\end{eqnarray}
and
\begin{eqnarray}
\nabla_{\alpha}\nabla_{\mu}(f_{X}R^{\alpha}_{\nu})+\nabla_{\alpha}\nabla_{\nu}(f_{X}R^{\alpha}_{\mu})&=&R^{\alpha}_{\nu}\nabla_{\alpha}\nabla_{\mu}f_{X}+R^{\alpha}_{\mu}\nabla_{\alpha}\nabla_{\nu}f_{X}+\frac{1}{2}(\nabla_{\mu}f_{X})(\nabla_{\nu}R)+\frac{1}{2}(\nabla_{\nu}f_{X})(\nabla_{\mu}R)\nonumber\\
&+&(\nabla_{\alpha}f_{X})(\nabla_{\mu}R^{\alpha}_{\nu})+(\nabla_{\alpha}f_{X})(\nabla_{\nu}R^{\alpha}_{\mu})+f_{X}\nabla_{\mu}\nabla_{\nu}R+2f_{X}R_{\alpha\mu\nu\lambda}R^{\alpha\lambda}\nonumber\\
&+&2f_{X}R_{\mu\lambda}R^{\lambda}_{\nu},
\end{eqnarray}
inserting them into Eq. (\ref{19}), the tress-energy tensor of the effective curvature fluid $\Omega_{\mu\nu}$ is simplified as
\begin{eqnarray}
\label{21}
\Omega_{\mu\nu}&=&\frac{1}{f_{R}}\Big[\frac{1}{2}g_{\mu\nu}(f-Rf_{R})+\nabla_{\mu}\nabla_{\nu}f_{R}-g_{\mu\nu}\Box f_{R}\nonumber\\
&-&f_{X}\Box R_{\mu\nu}-R_{\mu\nu}\Box f_{X}-g_{\mu\nu}R^{\alpha\beta}\nabla_{\alpha}\nabla_{\beta}f_{X}-\frac{1}{2}f_{X}g_{\mu\nu}\Box R\nonumber\\
&+&R^{\alpha}_{\nu}\nabla_{\alpha}\nabla_{\mu}f_{X}+R^{\alpha}_{\mu}\nabla_{\alpha}\nabla_{\nu}f_{X}-f_{X}\nabla_{\mu}\nabla_{\nu}R-2f_{X}R_{\alpha\mu\nu\lambda}R^{\alpha\lambda}\Big].
\end{eqnarray}
For a static spherically symmetric black hole whose geometry is given by
\begin{eqnarray}
\label{metric}
ds^2=-B(r)dt^2+\frac{dr^2}{B(r)}+r^2d\Omega^2,
\end{eqnarray}
where the event horizon is located at $r=r_+$ the largest positive root of $B(r_+)=0$ with $B'(r_+)\neq 0$, the $(^{1}_{1})$ components of the Einstein tensor is
\begin{eqnarray}
\label{22}
G^{1}_{1}=\frac{1}{r^{2}}\left(-1+rB^{\prime}+B\right),
\end{eqnarray}
with the primes denoting the derivative with respect to $r$. At the horizon, since $B(r_{+})=0$, it reduces to
\begin{eqnarray}
\label{23}
G^{1}_{1}=\frac{1}{r^{2}}\left(-1+rB^{\prime}\right).
\end{eqnarray}
While the radial components of the tress-energy tensor of the effective curvature fluid $\Omega_{\mu\nu}$ at the horizon takes the following form
\begin{eqnarray}
\label{24}
\Omega^{1}_{1}=\frac{1}{f_{R}}\left[\frac{1}{2}(f-Rf_{R})-\frac{1}{2}B'f'_{R}-B'(f_{X}R^{1}_{1})'-\frac{2f_{X}B'R^{1}_{1}}{r_{+}} +\frac{2f_{X}B'R^{3}_{3}}{r_{+}}-2f_{X}(R ^{1}_{1})^2\right].
\end{eqnarray}
Substituting Eqs. (\ref{23}) and (\ref{24}) into Eq. (\ref{18}), and thinking of $P=T^{r}_{~r}|_{r_{+}}$, we derive
\begin{eqnarray}
\label{25}
8\pi P=-\frac{f_{R}}{r^2_{+}}+\frac{f_{R}B'}{r_{+}}-\frac{1}{2}(f-Rf_{R})+\frac{1}{2}B'f'_{R}+B'(f_{X}R^1_1)'+\frac{2f_{X}B'R^1_1}{r_{+}}-\frac{2f_{X}B'R^3_3}{r_{+}}+2f_{X}(R^1_1)^2.
\end{eqnarray}
This equation is very complicated, how to determine the function $C(r_+)$ in Eq. (\ref{p1})? Even worse, this equation depends on higher derivatives of $B$, we can no longer do as what we have done in Einstein's theory in which we can obtain the entropy and energy from Eqs. (\ref{11}) and (\ref{energy}) directly. In higher-derivative gravity, to use the horizon first law, $\delta E=T\delta S-P\delta V$, usually one should reduce the higher-derivative field equations to lower-derivative field equations via a Legendre transformation \cite{Magnano:1987zz, Jakubiec:1988ef}. Here we try a new method. If we obtain the entropy by using other methods, then using the new horizon first law (\ref{nhfl}) and the degenerate Legendre transformation $E=G+TS-PV$, we can derive the energy. Since $f(R,R^{\mu\nu}R_{\mu\nu})$ is a diffeomorphism invariance of gravitational theory, the entropy can be obtained with Wald method, which is presented in the Appendix. Taking into account the volume of black hole $V(r_{+})=4\pi r^{3}_{+}/3$, the pressure in Eq. (\ref{25}), the Hawking temperature $T=B'(r_{+})/4\pi$, and the entropy given in the Appendix, the new horizon first law (\ref{nhfl}) can be rewritten as
\begin{eqnarray}
\label{g}
\delta G=-\frac{1}{3}\pi r^{2}_{+}(f_{R}+2f_{X}R^{1}_{1})\delta T+\frac{4\pi r^{3}_{+}T}{3}\delta\left(\frac{f_{R}}{2r_{+}}\right)+\frac{4\pi r^{3}_{+}T}{3}\delta\left(\frac{f_{X}R^{1}_{1}}{r_{+}}\right)+\frac{1}{3}\pi r^{3}_{+}\delta(Tf'_{R})+\frac{2}{3}\pi r^{3}_{+}\delta[T(f_{X}R^{1}_{1})']\nonumber\\
-\frac{1}{12}r^{3}_{+}\delta(f-Rf_{R})-\frac{1}{6}r^{3}_{+}\delta\left(\frac{f_{R}}{r^{2}_{+}}\right)+\frac{4\pi r^{3}_{+}}{3}\delta\left(\frac{4\pi f_{X}T^{2}}{r^{2}_{+}}\right)-\frac{4\pi r^{3}_{+}}{3}\delta\left(\frac{Tf_{X}}{r^{3}_{+}}\right)+\frac{1}{3}r^{3}_{+}\delta [f_{X}(R^{1}_{1})^{2}],
\end{eqnarray}
and $TS-PV$ is given by
\begin{eqnarray}
\label{tp}
TS-PV=\frac{1}{3}\pi r^{2}_{+}f_{R}T+\frac{2}{3}\pi r^{2}_{+}f_{x}R^{1}_{1}T-\frac{2}{3}\pi r^{3}_{+}T(f_{x}R^{1}_{1})'-\frac{1}{3}\pi r^{3}_{+}Tf'_{R}\nonumber\\+\frac{1}{6}f_{R}r_{+}+\frac{1}{12}r^{3}_{+}(f-Rf_{R})-\frac{1}{3}r_{+}f_{X}B'^{2}+\frac{1}{3}f_{X}B'-\frac{1}{3}r^{3}_{+}f_{X}(R^{1}_{1})^{2}.
\end{eqnarray}
According to the degenerate Legendre transformation $E=G+TS-PV$, we have
\begin{eqnarray}
\label{energy1}
\delta E=\delta (G+TS-PV)=\left[\frac{1}{2}f_{R}\delta r_{+}+\frac{1}{4}r^{2}_{+}(f-Rf_{R})\delta r_{+}+r_{+}f_{X}B'R^{3}_{3}\delta r_{+}-r^{2}_{+} f_{X}(R^{1}_{1})^{2}\right]\delta r_{+},
\end{eqnarray}
or equivalently
\begin{eqnarray}
\label{energy1}
E=\int^{r_{+}}\left[\frac{1}{2}f_{R}+\frac{1}{4}r^{2}_{+}(f-Rf_{R})+r_{+}f_{X}B'R^{3}_{3}-r^{2}_{+} f_{X}(R^{1}_{1})^{2}\right]dr_{+}.
\end{eqnarray}
When $f_{X}=0$, Eq. (\ref{energy1}) returns to the result obtained in $f(R)$ theory \cite{Zheng:2018fyn}. Using Eqs. (\ref{energy1}) and (\ref{e1}), we can calculate the energy and the entropy of black hole in $f(R,R^{\mu\nu}R_{\mu\nu})$ theory.
\section{Application: quadratic-curvature gravity}
For application, we consider a simple but important example: the most general quadratic-curvature gravity theory with a cosmological constant in four dimensions, its Lagrangian density is given by an arbitrary combination of scalar curvature-squared and Ricci-squared terms, namely
\begin{eqnarray}
\label{case1}
f(R,R^{\mu\nu}R_{\mu\nu})=R+\alpha R^{\mu\nu}R_{\mu\nu}+\lambda R^2-2\Lambda,
\end{eqnarray}
where $\alpha$ and $\lambda$ are constants, $\Lambda$ is the cosmological constant. For this theory, we have $f_{R}=1+2\lambda R$ and $f_{X}=\alpha$. After some calculations, we find from (\ref{e1}) that in spacetime with metric (\ref{metric}) the entropy is
\begin{eqnarray}
\label{s1}
S&=&\pi r^{2}_{+}(1+2\lambda R+2\alpha R^{1}_{1}).
\end{eqnarray}
Substituting Eq. (\ref{case1}) into Eq. (\ref{energy1}), the energy of the black hole is given by
\begin{eqnarray}
\label{e1}
E&=&\frac{1}{2}\int r^{2}_{+}\left[\frac{1}{r^2_{+}}+\frac{2\lambda R}{r^{2}_{+}}+\frac{1}{2}\alpha R^{\mu\nu}R_{\mu\nu}-\frac{1}{2}\lambda R^{2}+\frac{2\alpha B'R^{3}_{3}}{r_{+}}-2\alpha{R^{1}_{1}}^{2}-\Lambda\right]dr_+\\\nonumber
&=&-\frac{1}{4}\int\left[\left(6\alpha{R^{1}_{1}}^{2}+2R^{1}_{1}\alpha B''+2\alpha{R^{2}_{2}}^{2}+2\Lambda+\lambda R^{2}\right)r^{2}_{+}+4R^{1}_{1}\alpha r_{+}B'-2(2\alpha R^{2}_{2}+2\lambda R+1)\right]dr_+.
\end{eqnarray}
In the case of a Schwarzschild-(A)ds black hole, for example, whose metric is given by $B(r)=1-\frac{2M}{r}-\frac{\Lambda r^{2}}{3}$, the entropy (\ref{s1}) and the energy (\ref{e1}) respectively returns to
\begin{eqnarray}
\label{36}
S=\pi r^{2}_{+}(1+2\lambda R+2\alpha R^{1}_{1})=\pi r^{2}_{+}[1+2(\alpha+4\lambda)\Lambda],
\end{eqnarray}
and
\begin{eqnarray}
\label{38}
E=[1+2(\alpha+4\lambda)\Lambda]M,
\end{eqnarray}
where we have used $B(r_+)$=0. Eqs. (\ref{36}) and (\ref{38}) are consistent with the results obtained in \cite{Ma:2017jcy, Giribet:2018hck}. The energy in \cite{Giribet:2018hck} was computed by using Abbott-Deser-Tekin method and a qualitatively different way of regularizing the Iyer-Wald charges respectively. While in  \cite{Ma:2017jcy} it was calculated via the horizon first law after taking a Legendre transformation. When $\alpha=0$, we obtain the results in $R+\lambda R^2-2\Lambda$ theory. While for $\alpha=0$ and $\lambda=0$, we get the results in Einstein's gravity with cosmological constant.
\section{conclusion}
We have discussed whether the new horizons first law is still valid in $f(R, R^{\mu\nu}R_{\mu\nu})$ theory. Unlike done in Einstein's gravity, we can not directly derive the entropy and energy via Eqs. (\ref{11}) and (\ref{energy}). We must firstly obtain the entropy via other methods, such as Wald formula, then we can use the new horizon first law, the degenerate Legendre transformation, and the gravitational field equations under considering to derive the energy of black hole in $f(R,R^{\mu\nu}R_{\mu\nu})$ theory. For application, we have considered quadratic-curvature gravity and have presented the entropy and the energy for a static spherically symmetric black hole, especially for a Schwarzschild-(A)ds black hole where the results are consistent with those obtained in literatures. Whether this procedure can be applied to other complicated cases, such as Einstein-Horndeski-Maxwell theory \cite{Feng2016}, is worth future studying. It is also interesting to compare our results with those may be obtained by using other methods, such as Misner-Sharp and Abbott-Deser-Tekin methods, which will be investigated in our future studies.

\begin{acknowledgments}
We thank J. Zhai for helpful advice. This study is supported in part by Hebei Provincial Natural Science Foundation of China (Grant No. A2014201068).
\end{acknowledgments}

\section{Appendix}
In the bulk of the paper we studied the entropy and the energy of black hole (\ref{metric}) in $f(R,R^{\mu\nu}R_{\mu\nu})$ theory. To confirm the results are reasonable, here we calculate the entropy by using the Wald formula which takes the form
\begin{eqnarray}
\label{1}
S=-2\pi\oint\frac{\delta L}{\delta R_{abcd}} \epsilon_{ab}\epsilon_{cd}~dV_2
\end{eqnarray}
where $L$ is Lagrangian density of gravitational field, $dV_2$ is the volume element on the bifurcation surface $\Sigma$, and $\epsilon_{ab}$ is the binormal vector to $\Sigma$ normalized as $\epsilon_{ab}\epsilon^{ab}=-2$. For the metric (\ref{metric}) the binormal vectors can easily be found as $\epsilon_{01}=1$ and $\epsilon_{10}=-1$. For $L=\frac{f(R,R_{\mu\nu}R^{\mu\nu})}{16\pi}$, one straightly gets
\begin{eqnarray}
\label{2}
\frac{\delta L}{\delta R_{abcd}}&=&\frac{1}{16\pi}\left(f_{R}\frac{\delta R}{\delta R_{abcd}}+f_{X}\frac{\delta R_{\mu\nu}R^{\mu\nu} }{\delta R_{abcd}}\right)\\\nonumber
&=&\frac{1}{16\pi}\left(g^{c[a}g^{b]d}f_{R}+2R^{\mu\nu}g^{\sigma\rho}\delta^{a}_{[\mu}\delta^{b}_{\sigma]}\delta^{c}_{\nu}\delta^{d}_{\rho}f_{X}\right).
\end{eqnarray}
For the metric (\ref{metric}), we have $g^{c[a}g^{b]d}\epsilon_{ab}\epsilon_{cd}=-2$ and $
2R^{\mu\nu}g^{\sigma\rho}\delta^{a}_{[\mu}\delta^{b}_{\sigma]}\delta^{c}_{\nu}\delta^{d}_{\rho}\xi_{ab}\xi_{cd}=4R^{00}g^{11}=-4R^{1}_{1}$. Since the integral (\ref{1}) is to be evaluated on shell, finally we have the entropy
\begin{eqnarray}
\label{e1}
S=\frac{A({r_{+})}}{4}\left(f_{R}+2f_{X}R^{1}_{1}\right)
\end{eqnarray}
in $f(R,R^{\mu\nu}R_{\mu\nu})$ theory for the black hole (\ref{metric}).

\bibliographystyle{ieeetr}
\bibliography{ref}
\end{CJK*}
\end{document}